\documentclass{aa}

\usepackage{natbib}
\usepackage{amsmath}
\usepackage{amsfonts}
\usepackage{amssymb}
\usepackage{graphicx}
\usepackage[titletoc]{appendix}
\usepackage{color}
\usepackage{hyperref}
\usepackage{cleveref}
\usepackage[rightcaption]{sidecap}
\usepackage{subfigure}
\usepackage{comment}
\usepackage{soul}
\usepackage{cancel}
\usepackage{dcolumn}

\usepackage{array}
\usepackage{ctable}
\usepackage{multirow}
\usepackage{siunitx}
\usepackage{longtable}
\usepackage{tabularx}
\usepackage{booktabs}
\usepackage{float}

\graphicspath{{Graphics/}}

\def\be{\begin{equation}}
\def\ee{\end{equation}}
\def\bea{\begin{eqnarray}}
\def\eea{\end{eqnarray}}

\def\prd{Phys. Rev. D}
\def\mnras{MNRAS}
\def\aj{AJ}
\def\apj{ApJ}
\def\apjl{ApJ Lett.}
\def\apjs{ApJ Suppl. Ser.}

\def\aap{A\&A}

\def\araa{Annual Rev. of Astron. Astrophys.}

\def\jcap{JCAP}

\definecolor{vividviolet}{rgb}{0.62, 0.0, 1.0}
\definecolor{amaranth}{rgb}{0.9, 0.17, 0.31}
\definecolor{palatinateblue}{rgb}{0.15, 0.23, 0.89}
\definecolor{brightpink}{rgb}{1.0, 0.0, 0.5}
\definecolor{cornflowerblue}{rgb}{0.39, 0.58, 0.93}
\definecolor{deepcarminepink}{rgb}{0.94, 0.19, 0.22}
\definecolor{radicalred}{rgb}{1.0, 0.21, 0.37}

\hypersetup{ linktoc=all,
    colorlinks, linkcolor={palatinateblue},
    citecolor={brightpink}, urlcolor={amaranth}
}

\DeclareUnicodeCharacter{2212}{\ensuremath{-}}

\begin{document}

\title{Dark energy reconstructions combining BAO data with galaxy clusters and intermediate redshift catalogs}

\author{Orlando Luongo\inst{1,2,3,4,5}
\and
Marco Muccino\inst{1,5,6,7}
}

\institute{Universit\`a di Camerino, Divisione di Fisica, Via Madonna delle carceri 9, 62032 Camerino, Italy.
\and
SUNY Polytechnic Institute, 13502 Utica, New York, USA.
\and
INAF - Osservatorio Astronomico di Brera, Milano, Italy.
\and
INFN, Sezione di Perugia, Perugia, 06123, Italy.
\and
Al-Farabi Kazakh National University, Al-Farabi av. 71, 050040 Almaty, Kazakhstan.
\and
Institute of Nuclear Physics, Ibragimova, 1, 050032 Almaty, Kazakhstan.
\and
ICRANet, Piazza della Repubblica 10, Pescara, 65122, Italy.\\
\email{orlando.luongo@unicam.it,\,marco.muccino@lnf.infn.it}}

\date{}

  \abstract
   {Cosmological parameters and dark energy (DE) behavior are generally constrained assuming \textit{a priori} models.}
   {We work out a model-independent reconstruction to bound the key cosmological quantities and the DE evolution.}
   {Through the model-independent \textit{Bézier interpolation} method, we reconstruct the Hubble rate from the observational Hubble data and derive analytic expressions for the distances of galaxy clusters, type Ia supernovae, and uncorrelated baryonic acoustic oscillation (BAO) data. In view of the discrepancy between Sloan Digital Sky Survey (SDSS) and Dark Energy Spectroscopic Instrument (DESI) BAO data, they are kept separate in two distinct analyses. Correlated BAO data are employed to break the baryonic--dark matter degeneracy. All these interpolations enable us to single out and reconstruct the DE behavior with the redshift $z$ in a totally model-independent way.}
   {In both analyses, with SDSS-BAO or DESI-BAO data sets, the constraints agree at $1$--$\sigma$ confidence level (CL) with the flat $\Lambda$CDM model. The Hubble constant tension appears solved in favor of the Planck satellite value. The reconstructed DE behavior exhibits deviations at small $z$ ($>1$--$\sigma$ CL), but agrees ($<1$--$\sigma$ CL) with the cosmological constant paradigm at larger $z$.}
   {Our method hints for a slowly evolving DE, consistent with a cosmological constant at early times.}


\titlerunning{Dark energy reconstructions combining BAO with GCs and intermediate redshift catalogs}
\authorrunning{Luongo \& Muccino}

\maketitle

\section{Introduction}

Our current understanding of the Universe does not adequately explain the physical reasons behind its accelerated expansion, experimentally found by observations of SNe Ia \citep[see][]{1998AJ....116.1009R,1999ApJ...517..565P}. To account for this unexpected phenomenon, it is widely-accepted that an exotic fluid, exhibiting a negative equation of state, might be included into Einstein’s energy-momentum tensor.

Indeed, the presence of baryonic and cold dark matter alone is insufficient to describe the Universe's late-time acceleration \citep{2003RvMP...75..559P}, leading to the hypothesis of a time-dependent fluid, known as  DE, that drives the Universe to speed up.

Among various DE models, the $\Lambda$CDM paradigm posits that DE is in the form of a genuine cosmological constant \citep{2001LRR.....4....1C}. Thus, it is not hard to believe that this model, due to its minimal set of free parameters, is the most statistically preferred for describing large-scale cosmic dynamics, making it particularly suitable for late-time cosmology.
However, recent observational cosmological tensions, such as the Hubble constant tension and inconsistencies in the clustering amplitude $S_8$ across low and high-redshift measurements \citep{bernalcosmic, 2021CQGra..38o3001D, 2022JHEAp..34...49A}, as well as unresolved theoretical challenges like the coincidence and fine-tuning problems  \citep{2006IJMPD..15.1753C}, have motivated the exploration of alternative models \cite{2024arXiv240917019W,2024PhRvD.110h3528W,2023PhRvD.108j3519W}. New data, including DESI measurements of baryonic acoustic oscillations (BAO), support numerous approaches that may replicate or even surpass the predictive successes of the $\Lambda$CDM paradigm \citep{nostro,2022CQGra..39s5014D,mio2022}.

To mitigate reliance on a specific cosmological model, various model-independent methods have been proposed \citep{2013Galax...1..216C, 2016IJGMM..1330002D, 2021MNRAS.503.4581L, 2023MNRAS.518.2247L, 2007MNRAS.380.1573S, 2010PhRvD..81h3537S, 2018JCAP...10..015H}.
A primary challenge of these approaches is the difficulty of reconciling data from low, intermediate, and early cosmic epochs, aiming to capture DE’s evolution across the full expansion history. Additionally, a significant drawback arises from assuming both cold dark matter and baryons as dust-like matter, which prevents separate measurements of their respective densities, creating a degeneracy within the matter sector.

In this work, we propose a strategy that heals the above issues by combining different probes -- based on low, intermediate and early time data points -- and disentangles the matter sector in order to mold DE at different stages of its evolution.
We resort a model-independent approach based on the so-called B\'ezier parametric interpolation \citep{2019MNRAS.486L..46A,2021MNRAS.503.4581L,2021MNRAS.501.3515M,2023MNRAS.518.2247L,2023MNRAS.523.4938M,2024JHEAp..42..178A,2024A&A...686A..30A,2024arXiv240802536A}, which is used to:
\begin{itemize}
    \item[-] reconstruct the Hubble rate $H(z)$ fitting the observational Hubble data (OHD) \citep[see, e.g.,][]{2019MNRAS.486L..46A},
    \item[-] derive analytic expressions for the distances of galaxy clusters (GCs), SNe Ia, and BAO uncorrelated data,
    \item[-] break the baryonic--dark matter degeneracy within the comoving sound horizon $r_{\rm d}$ \citep{1999MNRAS.304...75E}, through the interpolation of the correlated BAO data.
\end{itemize}

To check whether our method can give additional insights on the form of dark energy,
\begin{itemize}
    \item[-] we seek non-flat Universe, adding the spatial curvature into the Friedmann equations,
    \item[-] we analyze the impact of our strategy on the cosmological tensions.
\end{itemize}

Results from our Monte Carlo–Markov chain (MCMC) simulations, using the Metropolis-Hastings algorithm, indicate that this approach provides valuable insights.
Precisely, when comparing our results with expectations from the flat (non-flat) $\Lambda$CDM model, the constraints agree within $1$--$\sigma$ CL with the flat $\Lambda$CDM model, whereas the Hubble constant tension is solved in favor of the Planck satellite value.
Accordingly, the reconstructed DE behavior exhibits deviations at small $z$ ($\gtrsim 1$--$\sigma$ CL), but agrees with the cosmological constant paradigm at larger $z$.
Consequently, our method suggests a slowly evolving DE, however consistent with a cosmological constant at early times.

The paper is structured as follows. The methods of our treatment are reported in Sect.~\ref{sec2}, where the B\'ezier interpolation is explained in detail. The numerical findings are thus reported in Sect.~\ref{sec3}, where the MCMC analyses are summarized. The core of DE reconstructions is displayed and theoretically discussed in Sect.~\ref{sec4}. Conclusions and perspectives are summarized in Sect.~\ref{sec5}.

\section{Methods}\label{sec2}

In this section, we describe the methodologies developed throughout our manuscript in order to obtain model-independent cosmological bounds. To do so,
\begin{itemize}
\item[-] we follow the methodology introduced in \citet{2024A&A...686A..30A} that makes use of OHD, GCs and BAO intermediate redshift catalogs,
\item[-] we include the \emph{Pantheon}+ catalog of SNe Ia \citep{2022ApJ...938..113S} to refine the overall constraints, and
\item[-] we perform two separate MCMC analyses, depending whether either SSDS or DESI data are involved into computation, in view of the claimed evidence for evolving DE derived from DESI-BAO data \citep{2024arXiv240403002D}.
\end{itemize}

Then, we jointly fit OHD, GC, SNe Ia, and BAO catalogs, based on the key steps, summarized below.
\begin{itemize}
\item[1)] The B\'ezier interpolation of the OHD catalog provides a model-independent expression for the Hubble rate $H(z)$ and an alternative estimate of the Hubble constant $H_0$.
\item[2)] This interpolation is used to derive analytic expressions for the angular diameter distance $D_{\rm A}(z)$ of GCs, the luminosity distance $D_{\rm L}(z)$ of SNe Ia, and BAO observables, which bear no \emph{a priori} assumptions on $\Omega_k$ that can be extracted from the fits.
\item[3)] The combination of SDSS-BAO or DESI-BAO data with correlated WiggleZ-BAO data \emph{breaks the baryonic--dark matter degeneracy} through the definition of $r_{\rm d}$.
\item[4)] The so-extracted cosmic bounds plus the $H(z)$ reconstruction single out and reconstruct DE behavior in terms of $z$, in a quite fully model-independent way.
\end{itemize}

The key feature of our recipe is therefore based on the use of B\'ezier approximation that we describe below.

\subsection{B\'ezier interpolation of $H(z)$}

The $N_{\rm O}=34$ OHD measurements of the Hubble rate (see Table~\ref{tab:OHD}) are obtained from the detection of couples of galaxies, assumed to form at same age, mostly and rapidly exhausted their gas reservoir, thence, evolving passively.
Once the difference in age, $\Delta t$, and redshift, $\Delta z$, of these pairs of galaxies are spectroscopically determined, the Hubble parameter is estimated from the identity $H(z)=-(1+z)^{-1}\Delta z/\Delta t$ \citep{2002ApJ...573...37J}.

For the sake of clearness, age-dating galaxies is affected by large systematic errors, typically associated with star formation history, stellar age, formation timescale, chemical composition, and so on. These uncertainties  contribute with additional $20$--$30\%$ errors \citep{2022LRR....25....6M}, leading to  measurements that are not particularly accurate. The great advantage relies on their determination though, i.e., it is roughly model-independent, as much as the above hypotheses on galaxy formation are fulfilled.

\begin{table}
\centering
\footnotesize
\setlength{\tabcolsep}{.7em}
\renewcommand{\arraystretch}{1.1}
   \begin{tabular}{lcl}
   \hline\hline
    $z$     &$H$ &  References \\
            &(km/s/Mpc)&\\
    \hline
    0.07  & $69.0  \pm 19.6\pm 12.4$ & \cite{Zhang2014} \\
    0.09    & $69.0  \pm 12.0\pm 11.4$  & \cite{Jimenez2002} \\
    0.12    & $68.6  \pm 26.2\pm 11.4$  & \cite{Zhang2014} \\
    0.17    & $83.0  \pm 8.0\pm 13.1$   & \cite{Simon2005} \\
    0.1791   & $75.0  \pm 3.8\pm 0.5$   & \cite{Moresco2012} \\
    0.1993   & $75.0  \pm 4.9\pm 0.6$   & \cite{Moresco2012} \\
    0.20    & $72.9  \pm 29.6\pm 11.5$  & \cite{Zhang2014} \\
    0.27    & $77.0  \pm 14.0\pm 12.1$  & \cite{Simon2005} \\
    0.28    & $88.8  \pm 36.6\pm 13.2$  & \cite{Zhang2014} \\
    0.3519   & $83.0  \pm 13.0\pm 4.8$  & \cite{Moresco2016} \\
    0.3802  & $83.0  \pm 4.3\pm 12.9$  & \cite{Moresco2016} \\
    0.4     & $95.0  \pm 17.0\pm 12.7$  & \cite{Simon2005} \\
    0.4004  & $77.0  \pm 2.1\pm 10.0$  & \cite{Moresco2016} \\
    0.4247  & $87.1  \pm 2.4\pm 11.0$  & \cite{Moresco2016} \\
    0.4497  & $92.8  \pm 4.5\pm 12.1$  & \cite{Moresco2016} \\
    0.47    & $89.0\pm 23.0\pm 44.0$     & \cite{2017MNRAS.467.3239R}\\
    0.4783  & $80.9  \pm 2.1\pm 8.8$   & \cite{Moresco2016} \\
    0.48    & $97.0  \pm 62.0\pm 12.7$  & \cite{Stern2010} \\
    0.5929   & $104.0 \pm 11.6\pm 4.5$  & \cite{Moresco2012} \\
    0.6797    & $92.0  \pm 6.4\pm 4.3$   & \cite{Moresco2012} \\
    0.75    & $98.8\pm33.6$     & \cite{2022ApJ...928L...4B}\\
    0.7812   & $105.0 \pm 9.4\pm 6.1$  & \cite{Moresco2012} \\
    0.80    & $113.1\pm15.1\pm 20.2$    & \cite{2023ApJS..265...48J}\\
    0.8754   & $125.0 \pm 15.3\pm 6.0$  & \cite{Moresco2012} \\
    0.88    & $90.0  \pm 40.0\pm 10.1$  & \cite{Stern2010} \\
    0.9     & $117.0 \pm 23.0\pm 13.1$  & \cite{Simon2005} \\
    1.037   & $154.0 \pm 13.6\pm 14.9$  & \cite{Moresco2012} \\
     1.26    & $135.0\pm 65.0$          & \cite{Tomasetti2023}\\
    1.3     & $168.0 \pm 17.0\pm 14.0$  & \cite{Simon2005} \\
    1.363   & $160.0 \pm 33.6$  & \cite{Moresco2015} \\
    1.43    & $177.0 \pm 18.0\pm 14.8$  & \cite{Simon2005} \\
    1.53    & $140.0 \pm 14.0\pm 11.7$  & \cite{Simon2005} \\
    1.75    & $202.0 \pm 40.0\pm 16.9$  & \cite{Simon2005} \\
    1.965   & $186.5 \pm 50.4$  & \cite{Moresco2015} \\
\hline
\end{tabular}
\caption{OHD catalog redshifts, values of $H$ with statistical and systematic (or combinrd) errors, and references, respectively.}
\label{tab:OHD}
\end{table}

The function of $z$ best-interpolating the OHD catalog is a second order B\'ezier curve
\begin{equation}
\label{bezier1}
\mathcal H(z) = \frac{\alpha_\star}{z_{\rm m}^2} \left[\alpha_0(z_{\rm m}-z)^2 + 2\alpha_1 z(z_{\rm m}-z) +\alpha_2 z^2\right]\,,
\end{equation}
with normalization $\alpha_\star=100\,{\rm km\,s}^{-1}{\rm Mpc}^{-1}$ and coefficients $\alpha_i$, that is extrapolated up to $z_{\rm m}=2.33$, which is the largest redshift for OHD, GC, Pantheon+, and BAO catalogs.
From Eq.~\eqref{bezier1}, at $z=0$ the dimensionless Hubble constant can be defined by $h_0\equiv H_0/\alpha_\star\equiv\alpha_0$.

With Gaussian distributed errors $\sigma_{H_j}$, the coefficients $\alpha_i$ are found by maximizing the log-likelihood function
\begin{equation}
\label{loglikeOHD}
    \ln \mathcal{L}_{\rm O} \propto -\frac{1}{2} \sum_{j=1}^{N_{\rm O}}\left[\dfrac{H_j-\mathcal H(z_j)}{\sigma_{H_j}}\right]^2\,.
\end{equation}

\subsection{Constraints on the curvature parameter from GC data}

When CMB photons travel across intra-cluster high-energy electrons in GCs, inverse Compton scattering occurs, causing the distortion of the CMB spectrum. This phenomenon is referred to as the Sunyaev-Zeldovich (SZ) effect \citep{1970CoASP...2...66S,1972CoASP...4..173S,2002ARA&A..40..643C}.

The SZ effect is redshift-independent and, combined with high signal-to-noise ratio X-ray surface brightness of the intra-cluster gas, which is redshift-dependent, it is possible to determine the triaxial structure of the GC and, thus, the corresponding corrected angular diameter distance $D_{\rm A}$.

Table~\ref{tab:SZ} lists the sample of $N_{\rm G}=25$ of such determined GC distances $D_{\rm A}$ \citep{2005ApJ...625..108D}.
The systematic errors -- mainly due to radio halos, X-ray absolute flux and electron temperature calibrations hindering the SZ effect calibration -- are typically $\approx13\%$ \citep{2006ApJ...647...25B}.

\begin{table}
\centering
\footnotesize
\setlength{\tabcolsep}{3.em}
\renewcommand{\arraystretch}{1.1}
\begin{tabular}{lc}
\hline\hline
$z$     & $D_{\rm A}$ \\
        & (Mpc)\\
\hline
$0.023$	& $103\pm42$\\
$0.058$	& $242\pm61$\\
$0.072$ & $165\pm45$\\
$0.074$ & $369\pm62$\\
$0.084$	& $749\pm385$\\
$0.088$	& $448\pm185$\\
$0.091$	& $335\pm70$\\
$0.142$	& $478\pm126$\\
$0.176$	& $809\pm263$\\
$0.182$	& $451\pm189$\\
$0.183$ & $604\pm84$\\
$0.202$	& $387\pm141$\\
$0.202$	& $806\pm163$\\
$0.217$	& $1465\pm407$\\
$0.224$	& $1118\pm283$\\
$0.252$	& $946\pm131$\\
$0.282$	& $1099\pm308$\\
$0.288$	& $934\pm331$\\
$0.322$	& $885\pm207$\\
$0.327$	& $697\pm183$\\
$0.375$	& $1231\pm441$\\
$0.451$	& $1166\pm262$\\
$0.541$	& $1635\pm391$\\
$0.550$	& $1073\pm238$\\
$0.784$	& $2479\pm1023$\\
\hline
\end{tabular}
\caption{GCs catalog from \citet{2005ApJ...625..108D} with redshift (first column) and diameter angular distances (second column).}
\label{tab:SZ}
\end{table}

Using Eq.~\eqref{bezier1}, we obtain an interpolated angular diameter distance defined as
\begin{equation}
\label{eq:da2}
\mathcal D_{\rm A}(z) = \frac{c\left(1+z\right)^{-1}}{\alpha_\star\alpha_0\sqrt{\Omega_k}} \sinh \left[\int_0^z \frac{\alpha_\star\alpha_0\sqrt{\Omega_k} dz^\prime}{\mathcal H(z^\prime)}\right]\ ,
\end{equation}
that holds for any value of the curvature parameter $\Omega_k$.

If the errors $\sigma_{D_{{\rm A}j}}$ are Gaussian distributed, $\alpha_i$ and $\Omega_k$ are obtained by maximizing the log-likelihood
\begin{equation}
\label{loglikeSZ}
    \ln \mathcal{L}_{\rm G} \propto -\frac{1}{2} \sum_{j=1}^{N_{\rm G}}\left[\dfrac{D_{{\rm A}j}-\mathcal D_{\rm A}(z_j)}{\sigma_{D_{{\rm A}j}}}\right]^2\,.
\end{equation}

\subsection{Reinforcing the constraints with Pantheon+ data}

The Pantheon+ is a catalog of $N_{\rm S}=1701$ SNe Ia with a redshift coverage $0<z\leq2.3$, that comprises 18 different samples \citep{2022ApJ...938..113S}.

The luminosity distance $D_{\rm L}$ (in Mpc) of each SN Ia with rest-frame $B$-band apparent magnitude $m$ is given by
\begin{equation}
\label{DLSN}
D_{\rm L} = 10^{\left(m - M - 25\right)/5}\,.
\end{equation}
The rest-frame $B$-band absolute magnitude $M$ in Eq.~\eqref{DLSN} can be viewed as a nuisance parameter. Following \citet{2011ApJS..192....1C}, the marginalization over $M$ that maximizes the log-likelihood $\ln{\mathcal{L}_{\rm S}}$ of SN Ia is $M = b/e$, leading to
\begin{equation}\label{eqn:chimarg}
 \ln{\mathcal{L}_{\rm S}} \propto -\frac{1}{2}\left[a + \ln\left(\frac{e}{2 \pi}\right) - \frac{b^2}{e}\right]\,,
\end{equation}
with $a\equiv\Delta\mathbf{m}^{T}\mathbf{C}^{-1}\Delta\mathbf{m}$, $b\equiv\Delta\mathbf{m}^{T}\mathbf{C}^{-1}\mathbf{1}$ and $e \equiv
\mathbf{1}^T\mathbf{C}^{-1} \mathbf{1}$.
In these definitions, $\mathbf{C}$ is the covariance matrix that includes statistical and systematic errors\footnote{\url{https://github.com/PantheonPlusSH0ES/DataRelease}}, and $\Delta \mathbf{m}\equiv \mathbf{m} - \mathbf{m}_{\rm th}(z)$ is the vector of residuals with respect to the model magnitude vector $\mathbf{m}_{\rm th}(z)$, with elements defined as
\begin{equation}
 m_{\rm th}(z) = 5\log\left[(1+z)^2\mathcal D_{\rm A}(z)\right] + 25\,,
\end{equation}
where $\mathcal D_{\rm A}(z)$ is given by Eq.~\eqref{eq:da2}.

\section{Breaking the baryon--dark matter degeneracy with BAO}

\begin{table*}
\centering
\footnotesize
\setlength{\tabcolsep}{.7em}
\renewcommand{\arraystretch}{1.2}
\begin{tabular}{llccccl}
\hline\hline
Survey          & $z$       & $D_{\rm M}/r_{\rm d}$
                            & $D_{\rm H}/r_{\rm d}$
                            & $D_{\rm V}/r_{\rm d}$
                            & $A$
                            &  References \\
\hline
6dFGS		    & $0.106$	& & & $2.98\pm0.13$ &
                            & \citet{2011MNRAS.416.3017B}\\
                \hline
SDSS MGS 		& $0.15$	& & & $4.51\pm0.14$ &
                            & \citet{2021PhRvD.103h3533A}\\
SDSS DR12		& $0.38$    & $10.27\pm0.15$
                            & $24.89\pm 0.58$ & &
                            & \citet{2021PhRvD.103h3533A}\\
SDSS DR12		& $0.51$    & $13.38\pm0.18$
                            & $22.43\pm 0.48$ & &
                            & \citet{2021PhRvD.103h3533A}\\
SDSS DR16 LRG	& $0.70$    & $17.65\pm0.30$
                            & $19.78\pm0.46$ & &
                            & \citet{2021PhRvD.103h3533A}\\
SDSS DR16 ELG	& $0.85$	& $19.50\pm1.00$
                            & $19.60\pm2.10$ & &
                            & \citet{2021PhRvD.103h3533A}\\
SDSS DR16 QSO	& $1.48$    & $30.21\pm0.79$
                            & $13.23\pm0.47$ & &
                            & \citet{2021PhRvD.103h3533A}\\
SDSS DR16 Ly$\alpha$--Ly$\alpha$
                & $2.33$    & $37.60\pm1.90$
                            & $8.93\pm0.28$ & &
                            & \citet{2021PhRvD.103h3533A}\\
SDSS DR16 Ly$\alpha$--QSO	
                & $2.33$    & $37.30\pm1.70$
                            & $9.08\pm0.34$ & &
                            & \citet{2021PhRvD.103h3533A}\\
                \hline
DESI BGS 		& $0.30$    & & & $7.93\pm0.15$ &
                            & \citet{2024arXiv240403002D}\\
DESI LRG1	 	& $0.51$ 	& $13.62\pm0.25$
                            & $20.98\pm0.61$ & &
                            & \citet{2024arXiv240403002D}\\
DESI LRG2       & $0.71$ 	& $16.85\pm0.32$
                            & $20.08\pm0.60 $& &
                            & \citet{2024arXiv240403002D}\\
DESI LRG+ELG    & $0.93$    & $21.71\pm0.28$
                            & $17.88\pm0.35$ & &
                            & \citet{2024arXiv240403002D}\\
DESI ELG        & $1.32$    & $27.79\pm0.69$
                            & $13.82\pm0.42$ & &
                            & \citet{2024arXiv240403002D}\\
DESI QSO		& $1.49$	& & & $26.07\pm0.67$ &
                            & \citet{2024arXiv240403002D}\\
DESI Ly$\alpha$--QSO
                & $2.33$ 	& $39.71\pm0.94$
                            & $8.52\pm0.17$ & &
                            & \citet{2024arXiv240403002D}\\
\hline
WiggleZ			& $0.44$  	& & & & $0.474\pm0.034$
                            & \citet{2012MNRAS.425..405B}\\
WiggleZ			& $0.6$ 	& & & & $0.442\pm0.020$
                            & \citet{2012MNRAS.425..405B}\\
WiggleZ			& $0.73$ 	& & & & $0.424\pm0.021$
                            & \citet{2012MNRAS.425..405B}\\
\hline
\end{tabular}
\caption{BAO catalogs with surveys (first column), redshifts (second column), measurements with errors (third--sixth columns), and references (last column). For BAO from SDSS, see also \url{https://www.sdss4.org/science/final-bao-and-rsd-measurements/}.}
\label{tab:BAO}
\end{table*}

BAO are density fluctuations of the baryonic matter, generated by acoustic density waves in the primordial Universe \citep{2008cosm.book.....W}.
Their characteristic scale, embedded in the galaxy distribution \citep{2019JCAP...10..044C}, corresponds to the maximum distance $r_\mathrm{d}$ covered by the acoustic waves before their ``froze in'' due to the decoupling of baryons.

Based on non-parametric reconstructions, \citet{2021PhRvD.104d3521A} proposed for $r_\mathrm{d}$ a very accurate expression
\begin{equation}
\label{eq:neutrino}
r_\mathrm{d} = \frac{a_1~e^{a_2\left(a_3+\omega_{\nu}\right)^2}}{a_4~\omega_b^{a_5}+a_6~\omega_m^{a_7}+a_8\left(\omega_b\hspace{1mm}\omega_m\right)^{a_9}}~ \mathrm{Mpc}\,,
\end{equation}
where the density parameter for massive neutrino species is fixed to
$\omega_{\nu}=0.000645$ \citep{2015PhRvD..92l3516A}, and the density parameters for baryons only $\omega_b=h_0^2\Omega_b$ and for baryonic + dark matter $\omega_m=h_0^2\Omega_m$ are the free parameters.
The numerical coefficients $a_i$ have values
\begin{equation}
\nonumber
\begin{array}{lll}
a_1 = 0.0034917, & a_2=-19.972694, & a_3=0.000336186,\\
a_4 = 0.0000305, & a_5=0.22752,    & a_6=0.0000314257,\\
a_7 = 0.5453798, & a_8=374.14994,  & a_9=4.022356899.\\
\end{array}
\end{equation}

Table~\ref{tab:BAO} lists the four kind of BAO measurements collected from four different surveys:
\begin{itemize}
\item[-] 6dF Galaxy Survey (6dFGS), that mapped the nearby Universe over nearly half the sky \citep{2011MNRAS.416.3017B};
\item[-] SDSS, providing galaxy and quasar spectroscopic surveys \citep{2021PhRvD.103h3533A};
\item[-] DESI, that collected galaxy and quasar optical spectra to measure DE effect \citep{2024arXiv240403002D};
\item[-] WiggleZ Dark Energy Survey, furnishing
correlated estimates of the acoustic parameter \citep{2012MNRAS.425..405B}.
\end{itemize}
BAO measurements are affected by systematics errors related to photometry or spectroscopy, survey geometries and discrete volumes, etc. that are below $0.5\%$ \citep{2021MNRAS.503.3510G,2024arXiv240403002D}.

Resorting Eqs.~\eqref{bezier1}, \eqref{eq:da2}, \eqref{DLSN} and \eqref{eq:neutrino}, BAO uncorrelated observables $X=\{D_{\rm M}/r_{\rm d}\,,D_{\rm H}/r_{\rm d}\,,D_{\rm V}/r_{\rm d}\}$ -- the transverse comoving distance, the Hubble rate distance, and the angle-averaged distance ratios with $r_{\rm d}$, respectively -- listed in Table~\ref{tab:BAO}, can interpolated by the quantities $\mathcal X = \{\mathcal X_1\,,\mathcal X_2\,,\mathcal X_3\}$, respectively, given by the following expressions:
\begin{subequations}
\begin{align}
    \label{DMrd}
    \mathcal X_1(z) &=\frac{(1+z) \mathcal D_{\rm A}(z)}{r_{\rm d}}\,,\\
    \label{DHrd}
    \mathcal X_2(z) &=\frac{c}{r_{\rm d} \mathcal H(z)}\,,\\
    \label{DVrd}
    \mathcal X_3(z) &= \left[z \mathcal X_1(z) \mathcal X_2^2(z)\right]^{1/3}\,.
\end{align}
\end{subequations}
Eqs.~\eqref{DMrd}--\eqref{DVrd} reinforce the constraints on $h_0$ and $\Omega_k$ and set bounds on $\omega_b$ and $\omega_m$ via Eq.~\eqref{eq:neutrino} that, however, introduces a degeneracy between $\omega_b$ and $\omega_m$ \citep{1999MNRAS.304...75E} which is generally broken by fixing $\omega_b$ with the value got from the CMB \citep{Planck2018} or Big Bang nucleosynthesis theory \citep{2024JCAP...06..006S}.

To break this degeneracy, we resort the correlated BAO acoustic parameter $A$ listed in Table~\ref{tab:BAO} \citep{2012MNRAS.425..405B}, which described by the interpolation
\begin{equation}
    \label{A}
    \mathcal A(z) = g_\star\sqrt{\omega_m}\left[\frac{(1+z)^2 \mathcal D_{\rm A}^2(z)}{c^2 z^2\mathcal H(z)}\right]^{1/3}\,,
\end{equation}
that does not depend upon $r_{\rm d}$ and hence enables constraints on $h_0$, $\Omega_k$ and only $\omega_m$ \citep{2024A&A...686A..30A}.

The log-likelihood function of each of the uncorrelated BAO data, with corresponding errors $\sigma_X$, is given by
\begin{equation}
\label{loglikeBAOu}
    \ln \mathcal{L}_{\rm X} \propto -\frac{1}{2} \sum_{j=1}^{N_{\rm X}} \left[\dfrac{X_j-\mathcal X(z_j)}{\sigma_{X_j}}\right]^2\,,
\end{equation}
whereas the log-likelihood function for correlated BAO data with covariance matrix $\mathbf{C}_{\rm B}$ \citep{2012MNRAS.425..405B} is
\begin{equation}
\label{loglikeBAOc}
\ln \mathcal{L}_{\rm A} \propto -\frac{1}{2} \Delta{\bf A}^{\rm T} \mathbf{C}_{\rm B}^{-1}
\Delta{\bf A}\,,
\end{equation}
with $\Delta{\bf A}\equiv A_j-\mathcal A(z_j)$. Combining Eqs.~\eqref{loglikeBAOu}--\eqref{loglikeBAOc} leads to the total BAO log-likelihood function
\begin{equation}
\label{loglikeBAO}
    \ln \mathcal{L}_{\rm B} = \sum_X\ln \mathcal{L}_{\rm X} + \ln \mathcal{L}_{\rm A}\,.
\end{equation}

\section{Numerical results}\label{sec3}

Before proceeding with the numerical analysis, it is worth comparing the BAO data argued either from SDSS or from DESI.

As pointed out by the \citet{2024arXiv240403002D}, the region of the sky and the redshift ranges (see Table~\ref{tab:BAO}) observed by DESI partially overlaps with those from the SDSS, therefore, a joint fit would require the knowledge of the covariance matrix.

To this end, Ref. \citet{2024arXiv240403002D} highlighted a large discrepancy  ($\sim 3$--$\sigma$) between the DESI and SDSS results, emphasized at redshift $z\sim0.7$.

In addition to the above considerations, recent works also evidenced possible anomalies in the DESI data set \citep{2024arXiv240408633C,2024A&A...690A..40L} and inconclusive evidence in favor of a dynamical DE over the standard cosmological paradigm \citep[see, e.g.,][for an overview]{2024arXiv240412068C,2024JCAP...10..035G,2024arXiv240413833W}.

For these reasons, DESI and SDSS data (see Table~\ref{tab:BAO}) will not be jointly fit, but rather will be kept separated into two MCMC analyses involving OHD, GCs, SNe Ia and the following combinations of BAO, dubbed as follows:
\begin{itemize}
    \item[-] {\bf MCMC1}, with BAO log-likelihood $\mathcal L_{\rm B1}$ given by Eq.~\eqref{loglikeBAO}, that combines the only data point from 6dFGS and $N_{\rm A}=3$ correlated data from WiggleZ with the $N_X=15$ uncorrelated measurements $X$ from SDSS;
    \item[-] {\bf MCMC2}, with BAO log-likelihood $\mathcal L_{\rm B2}$ given by Eq.~\eqref{loglikeBAO}, in which the data points from 6dFGS and WiggleZ are combined with $N_X=12$ uncorrelated measurements $X$ from DESI.
\end{itemize}

We get the best-fit parameters of MCMC1 and MCMC2 analyses by maximizing the log-likelihood functions
\begin{subequations}
\begin{align}
{\rm MCMC1:}\quad \ln{\mathcal{L}_1} &= \ln{\mathcal{L}_{\rm O}} + \ln{\mathcal{L}_{\rm G}} + \ln{\mathcal{L}_{\rm S}} + \ln{\mathcal{L}_{\rm B1}}\,,\\
{\rm MCMC1:}\quad \ln{\mathcal{L}_2} &= \ln{\mathcal{L}_{\rm O}} + \ln{\mathcal{L}_{\rm G}} + \ln{\mathcal{L}_{\rm S}} + \ln{\mathcal{L}_{\rm B2}}\,.
\end{align}
\end{subequations}
For both analyses, we impose a wide range of priors on the parameters of our model-independent reconstructions:
\begin{equation}
\nonumber
\begin{array}{rclcrclr}
\alpha_0\equiv h_0 & \in & \left[0,1\right], &\qquad & \qquad \Omega_k & \in &\left[-2,2\right],\\
\alpha_1 & \in & \left[0,2\right], & \quad &
\omega_b & \in & \left[0,1\right],\\
\alpha_2 & \in & \left[0,3\right], &\qquad \quad& \omega_m & \in & \left[0,1\right].
\end{array}
\end{equation}
Details on the MCMC1 and MCMC2 analyses and their corresponding $1$--$\sigma$ and $2$--$\sigma$ contour plots (see Figs.~\ref{fig:cont1}--\ref{fig:cont2}, respectively) can be found in Appendix~\ref{app:A}.

The best-fit values are listed in Table~\ref{tab:bestfit} and compared with the constrains on flat and non-flat $\Lambda$CDM model got from Planck TT, TE, EE+lowE+lensing data \citep{Planck2018}. In particular, in Fig.~\ref{fig:Bez} the flat $\Lambda$CDM case is used as a benchmark for the best-fitting B\'ezier interpolations of OHD, GC, SN Ia and BAO catalogs.

Focusing on the cosmological parameters $\omega_b$, $\omega_m$, $h_0$, and $\Omega_k$ (see Table~\ref{tab:bestfit}), we can deduce what follows below.
\begin{itemize}
\item[-] The MCMC1 analysis confirms and further refines the findings of \citet{2024A&A...686A..30A}, which where based on SDSS data points though not in their final version presented by \citet{2021PhRvD.103h3533A}.
\item[-] The MCMC2 bounds tend to agree with the MCMC1 results, albeit with a) a smaller value of $\omega_b$ and b) a barely-consistent (at $\approx1$--$\sigma$ CL) and positive $\Omega_k$.
\item[-] In both the analyses, the inclusion of the Pantheon+ catalog improved the constraints on $\Omega_k$, which are more compatible (within $1$--$\sigma$ CL) with the flat scenario or with small spatial curvature geometries.
\item[-] Both MCMC1 and MCMC2 results are in agreement within $1$--$\sigma$ CL with the flat concordance model, though with larger attached errors.
\item[-] For both analyses the consistency with the non-flat extension of the $\Lambda$CDM \citep{Planck2018} is at $2$--$\sigma$ CL, due to the estimate on $h_0$.
\item[-] The Hubble tension seems to be solved in favor of the \citet{Planck2018} value $h_0 = 0.6736\pm 0.0054$, consistent at $1$--$\sigma$ CL with both MCMC1 and MCMC2 estimates, whereas the value $h_0=0.7304\pm0.0104$ got from SNe Ia \citep{2022ApJ...934L...7R} is only consistent within $2$--$\sigma$ CL with MCMC1 and MCMC2 analyses.
\item[-] In general, the best-fit values got from the MCMC2 analysis (performed using the DESI-BAO data) seems to be closer to the flat $\Lambda$CDM best-fits \citep{Planck2018} than those from the MCMC1 procedure.
\end{itemize}

\begin{table*}
\centering
\footnotesize
\setlength{\tabcolsep}{0.3em}
\renewcommand{\arraystretch}{1.3}
\begin{tabular}{lllllll}
\hline\hline
$\alpha_0\equiv h_0$            &
$\alpha_1$                      &
$\alpha_2$                      &
$\Omega_k$                      &
$\omega_b$                      &
$\omega_m$                      &
$M$ (mag)                       \\
\hline
\multicolumn{7}{c}{MCMC1 analysis}\\
$0.682_{-0.027\,(0.041)}^{+0.023\,(0.039)}$ &
$1.169_{-0.055\,(0.085)}^{+0.053\,(0.086)}$ &
$2.33_{-0.11\,(0.18)}^{+0.11\,(0.19)}$      &
$-0.09_{-0.13\,(0.21)}^{+0.14\,(0.24)}$     &
$0.0253_{-0.0070\,(0.0100)}^{+0.0061\,(0.0109)}$ &
$0.145_{-0.018\,(0.029)}^{+0.022\,(0.035)}$  &
$-19.389_{-0.055\,(0.104)}^{+0.045\,(0.093)}$\\
\hline
\multicolumn{7}{c}{MCMC2 analysis}\\
$0.677_{-0.024\,(0.042)}^{+0.024\,(0.039)}$ &
$1.170_{-0.055\,(0.084)}^{+0.051\,(0.082)}$ &
$2.39_{-0.12\,(0.18)}^{+0.10\,(0.17)}$      &
$+0.08_{-0.12\,(0.19)}^{+0.13\,(0.21)}$     &
$0.0212_{-0.0051\,(0.0081)}^{+0.0075\,(0.0119)}$ &
$0.144_{-0.018\,(0.028)}^{+0.021\,(0.034)}$ &
$-19.407_{-0.049\,(0.099)}^{+0.051\,(0.099)}$\\
\hline
\multicolumn{7}{c}{Planck Collaboration}\\
$0.6736^{+0.0054}_{-0.0054}$    &
--                              &
--                              &
--                              &
$ 0.02237^{+0.00015}_{-0.00015}$&
$0.1430^{+0.0011}_{-0.0011}$    &
--                              \\
$0.636^{+0.021}_{-0.023}$       &
--                              &
--                              &
$-0.011^{+0.013}_{-0.012}$      &
$0.02249^{+0.00016}_{-0.00016}$ &
$0.1410^{+0.0015}_{-0.0015}$    &
--                              \\
\hline
\end{tabular}
\caption{MCMC best-fits and $1$--$\sigma$ ($2$--$\sigma$) errors, compared with flat and non-flat $\Lambda$CDM models \citep{Planck2018}.
}
\label{tab:bestfit}
\end{table*}
\begin{figure*}
\centering
\includegraphics[width=0.48\hsize,clip]{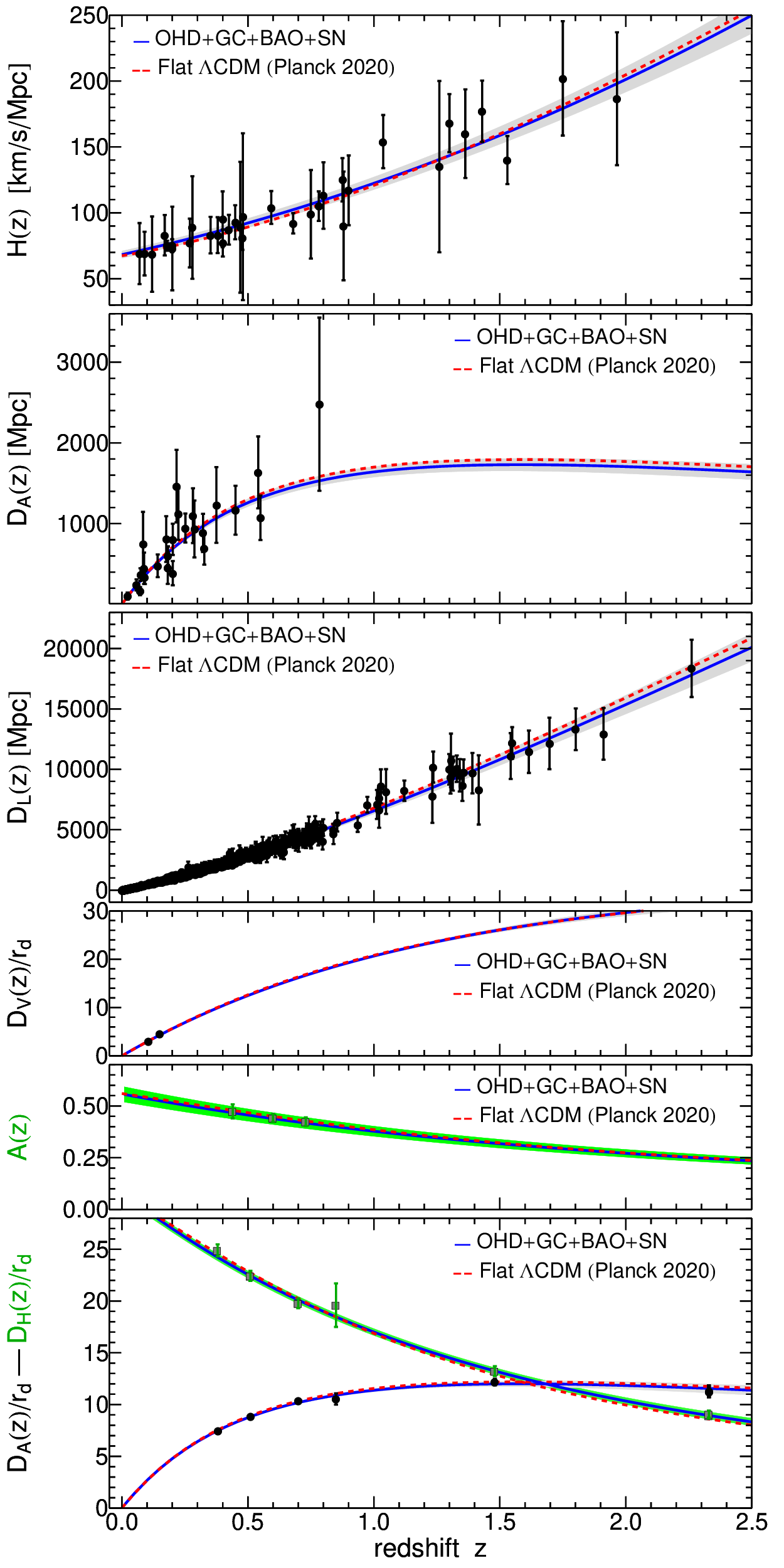}\hfill
\includegraphics[width=0.48\hsize,clip]{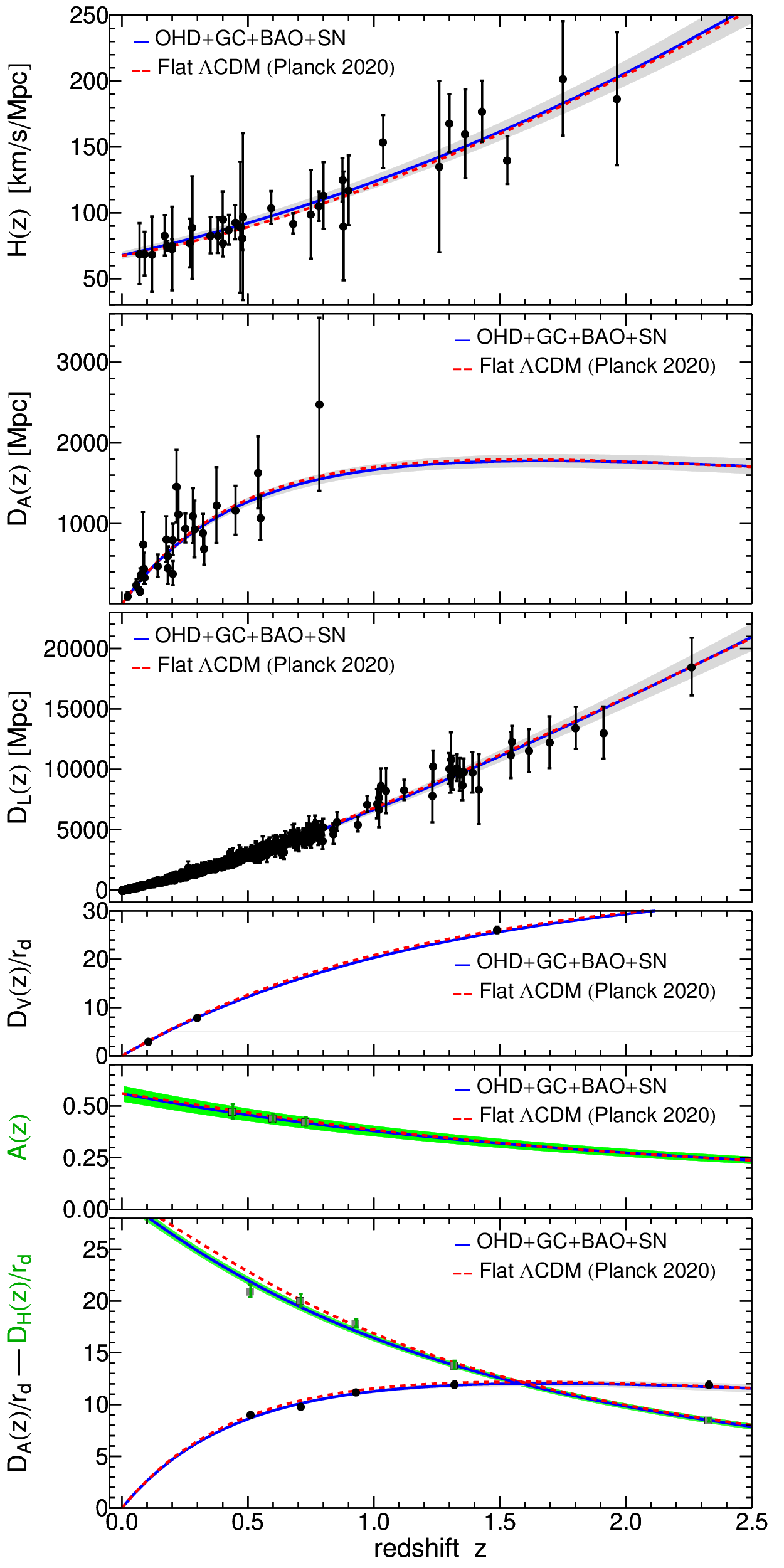}
\caption{Plots of OHD, GC, SN Ia and BAO data sets with MCMC1 (left column) and MCMC2 (right) best-interpolating B\'ezier curves (blue curves) and $1$--$\sigma$ confidence bands for $\mathcal H(z)$, $\mathcal D_{\rm A}(z)$, $(1+z)^2\mathcal D_{\rm A}(z)$, $\mathcal X_3(z)$, and $\mathcal X_1(z)/(1+z)$ (gray bands), and $\mathcal A(z)$ and $\mathcal X_2(z)$ (green bands), compared to the $\Lambda$CDM paradigm from the \citet{Planck2018} (dashed red curves).}
\label{fig:Bez}
\end{figure*}

\section{Reconstruction of the dark energy behavior}
\label{sec4}

From the best-fit values of Table~\ref{tab:bestfit} we can now attempt the reconstruction of the DE behavior.

We can use directly $\Omega_k$ to model the curvature contribution and the combination of $h_0$ and $\omega_m$ to constrain the matter density parameter.
Next, we use the CMB temperature $T_0 = 2.7255\pm 0.0006K$ and the effective extra relativistic degrees of freedom $N_{\rm eff}=2.99\pm0.17$ \citep{Planck2018} to compute the radiation density parameter $\Omega_r=9.15^{+0.26}_{-0.26}\times10^{-5}$.
Putting all these contributions together in a $\Lambda$CDM-like fashion, we define the following function of the redshift
\begin{equation}
\label{LCDMlike}
 f(z) = \alpha_0^{-2}\omega_m(1+z)^3 + \Omega_k (1+z)^2 + \Omega_r (1+z)^4\,.
\end{equation}
At this point, we can single out the contribution of the DE density by subtracting Eq.~\eqref{LCDMlike} from $\mathcal H^2(z)$, obtained by squaring Eq.~\eqref{bezier1}, namely
\begin{equation}
\label{DEz}
    \Omega_{de}(z) = \left(g_\star \alpha_0\right)^{-2} \mathcal H^2(z) - f(z) = \sum_{i=0}^{4} \beta_i (1+z)^i\,,
\end{equation}
where the coefficients $\beta_i$ in the last expressions and listed in Table~\ref{tab:bestfitbeta} depend upon combinations of the coefficients $\alpha_i$ and the cosmological parameters $\omega_m$, $\Omega_k$ and $\Omega_r$, as reported in Appendix~\ref{app:B}.
\begin{table*}
\centering
\footnotesize
\setlength{\tabcolsep}{1.6em}
\renewcommand{\arraystretch}{1.4}
\begin{tabular}{lccccc}
\hline\hline
                                &
$\beta_0$                       &
$\beta_1$                       &
$\beta_2$                       &
$\beta_3$                       &
$\beta_4$                       \\
\hline
MCMC1                           &
$0.324 \pm 0.068$               &
$0.284 \pm 0.062$               &
$0.355 \pm 0.091$               &
$-0.221 \pm 0.028$              &
$0.0327 \pm 0.0083$             \\
MCMC2                           &
$0.328 \pm 0.066$               &
$0.262 \pm 0.062$               &
$0.197 \pm 0.086$               &
$-0.222 \pm 0.030$              &
$0.0392 \pm 0.0084$             \\
\hline
\end{tabular}
\caption{DE reconstruction best-fit coefficients $\beta_i$ and $1$--$\sigma$ errors for MCMC1 and MCMC2 analyses.}
\label{tab:bestfitbeta}
\end{table*}
\begin{figure*}
\centering
\includegraphics[width=0.49\hsize,clip]{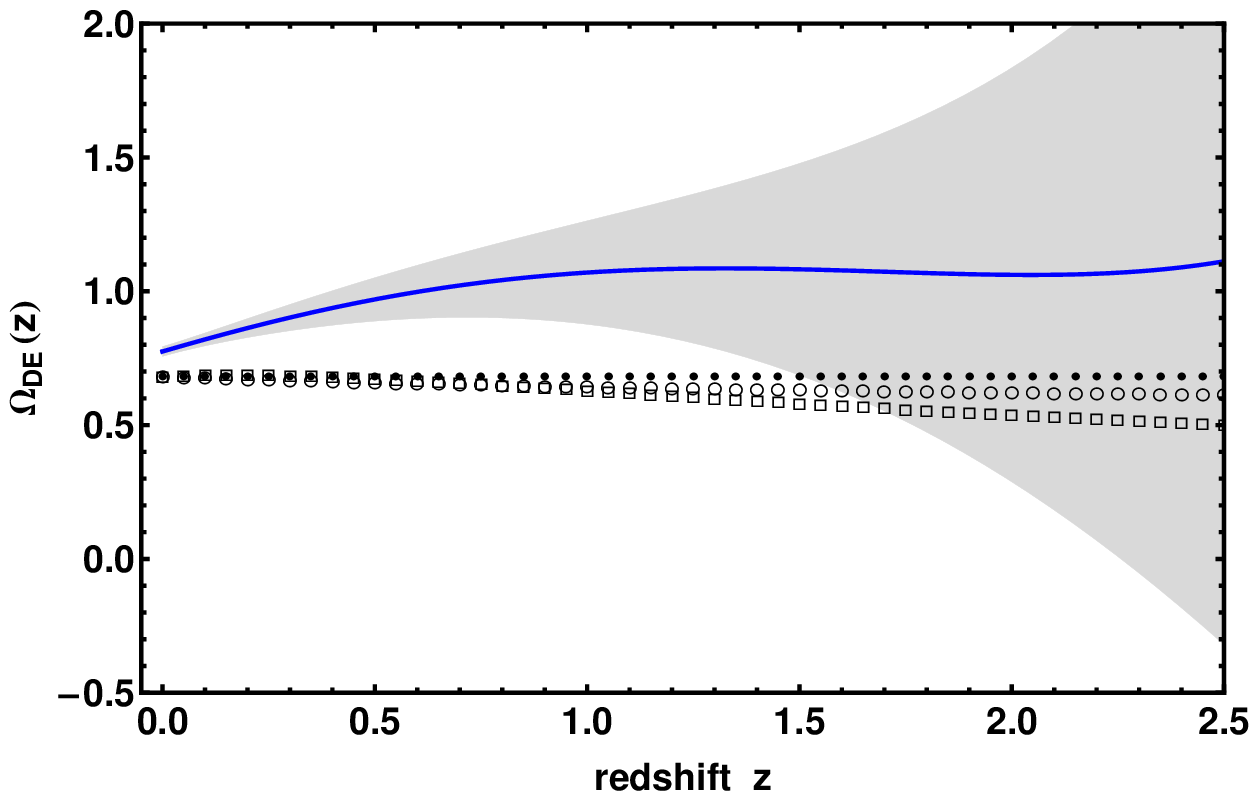}
\includegraphics[width=0.50\hsize,clip]{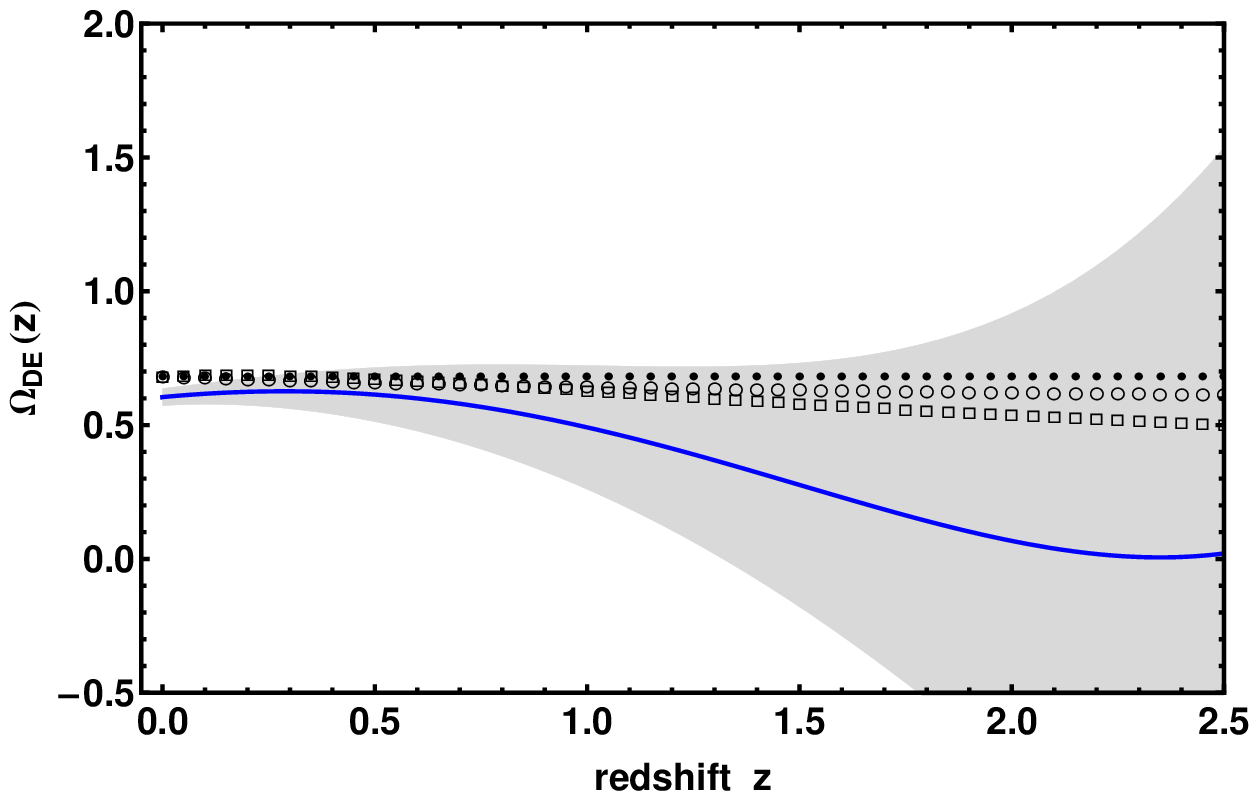}
\caption{Reconstructed DE behavior with $z$ (blue curve) with attached confidence band (gray area), compared to the expected behaviors for the $\Lambda$CDM (filled circles), the $w$CDM (empty circles) and the CPL (empty squares) models.}
\label{fig:Bez}
\end{figure*}

The DE reconstructed behaviors for both MCMC1 and MCMC2 analyses, obtained inputing the best-fit values of Table~\ref{tab:bestfit} in Eq.~\eqref{DEz}, are portrayed in Fig.~\ref{fig:Bez} and compared with the expectations for $\Lambda$CDM, $\omega$CDM and CPL models \citep{Planck2018}.

The behavior from the MCMC1 analysis indicates a $>1$--$\sigma$ deviation from the cosmological constant paradigm at $z\lesssim1.5$, with a preference for a dynamical DE behavior which differs (again, at $>1$--$\sigma$ CL) also from the simplest models, i.e., $\omega$CDM and CPL.

The MCMC2 analysis provides a reconstruction that, besides a small deviation at $z\lesssim0.3$, is always consistent (within $1$--$\sigma$ CL) with $\Lambda$CDM, $\omega$CDM and CPL models. This indicates that the DE equation of state slowly evolves with $z$, mimicking the behavior of the cosmological constant at $z\gtrsim0.3$ and deviating from it at smaller redshifts.

\section{Final outlooks and perspectives}
\label{sec5}

In this paper, we proposed a strategy to separately measure baryons and cold dark matter, by using different probes that may be classified in terms of low, intermediate and early redshift points. Disentangling
the matter sector we infer DE at different stages
of its evolution, resorting a model-independent approach, based on the use of B\'ezier interpolation.

In this respect, we extended the analysis performed in \citet{2024A&A...686A..30A} including the Pantheon+ catalog of SNe Ia, involving \emph{de facto} all low- and intermediate-redshift probes, above mentioned.

Further, in view of the recent BAO data release by the \citet{2024arXiv240403002D} -- showcasing a $\sim 3$--$\sigma$ discrepancy at $z\sim0.7$ and a claimed evidence for
dynamical DE with respect to the SDSS results -- we performed two separate MCMC analyses, namely

\begin{itemize}
    \item[-] MCMC1, involving SDSS data in conjunction with the other catalogs, and
    \item[-] MCMC2, replacing SDSS data set with DESI release.
\end{itemize}

Thus, the use of \emph{B\'ezier parametric curves} enables one to obtain model-independent bounds on $h_0$, $\Omega_k$, $\omega_b$ and $\omega_m$, having been used, in particular, to:

\begin{itemize}
    \item[(a)] infer an interpolated Hubble rate $\mathcal H(z)$ and estimate $h_0$ through OHD,
    \item[(b)] evaluate an analytic expressions for the angular diameter distance $\mathcal D_{\rm A}(z)$ of GCs, the luminosity distance $(1+z)^2 \mathcal D_{\rm A}(z)$ of SNe Ia, and BAO uncorrelated observables $\mathcal X$ with no \emph{a priori} assumptions on $\Omega_k$,
    \item[(c)] break the degeneracy between $\omega_b$ and $\omega_m$ in the definition of $r_{\rm d}$ with the interpolation of the correlated acoustic parameter $\mathcal A(z)$.
\end{itemize}

With respect to \citet{2024A&A...686A..30A}, the inclusion of the Pantheon+ catalog significantly improved the constraints on $\Omega_k$, further narrowing down the magnitude of a possible spatial curvature towards negligible values, and contributed in solving the Hubble tension in favor of CMB-consistent values, i.e., $h_0=0.682_{-0.027}^{+0.023}$ for MCMC1 and $h_0=0.677_{-0.024}^{+0.024}$ for MCMC2.
Notably, the inclusion of SNe Ia disfavors the estimate $h_0=0.7304\pm0.0104$ based on local SNe Ia anchored to Cepheid stars \citep{2022ApJ...934L...7R}.

In general, the cosmic bounds got from MCMC1 and MCMC2 (see Table~\ref{tab:bestfit} and Figs.~\ref{fig:cont1}--\ref{fig:cont2}) are in great agreement (within $1$--$\sigma$ CL) with the flat $\Lambda$CDM model.
In particular, albeit a slightly positive and small curvature, the MCMC2 analysis closely matches the expectations of the $\Lambda$CDM model \citep{Planck2018}, which is quite unexpected since DESI-BAO findings seem to question the standard paradigm \citep{2024arXiv240403002D}.

The corresponding results can be therefore summarized for the underlying analyses, namely:

\begin{itemize}
    \item[-] MCMC1, i.e., using SDSS-BAO data, highlighted a deviation from the cosmological constant at $z\lesssim1.5$ and an overall agreement at larger redshifts.
    \item[-] MCMC2, i.e., involving DESI-BAO data, exhibits a small deviation at $z\lesssim0.3$ and either an overall consistency (within $1$--$\sigma$ CL) with the $\Lambda$CDM paradigm or in line with a slowly evolving DE equation of state.
\end{itemize}

The latter appeared quite unexpectedly but in line with the above bounds on $h_0$, $\Omega_k$, $\omega_b$ and $\omega_m$.

Thus, the above improved cosmic bounds not only enabled a better reconstruction of $\mathcal H(z)$, but also the first attempt of a fully model-independent reconstruction of the DE evolution with $z$, based on the \emph{B\'ezier interpolation technique} that can be generalized in future works in terms of splines or alternatives to numerical reconstructions. Additionally, a further refinement in our model-independent procedure is mandatory, especially for the reconstruction and the modeling of DE equation of state and, therefore, for the understanding of its nature.
In this sense, besides improving all the hereby catalogs, it would be crucial for future works to:
\begin{itemize}
\item resolve the discrepancy between DESI and SDSS data sets and get joint cosmic bounds;
\item work out model- and probe-independent calibration methods for gamma-ray bursts \citep{2021Galax...9...77L} and quasars \citep{2019NatAs...3..272R}, investigating and  strengthening the constraints up to $z\sim9$.
\end{itemize}

Finally, the same procedure will be developed to additional DE scenarios to check the goodness of alternative models describing the cosmic speed up.

\section*{Acknowledgements}
OL acknowledges Gianluca Castignani for fruitful debates on the topic of this work during his stay at the University of Camerino. MM is grateful for the hospitality of the University of Camerino during the period in which this work has been written and acknowledges Grant No. BR21881941 from the Science Committee of the Ministry of Science and Higher Education of the Republic of Kazakhstan. The authors are thankful to Anna Chiara Alfano for interesting discussions related to the subject of model-independent techniques.

\bibliographystyle{aa}

\begin{thebibliography}{70}
\expandafter\ifx\csname natexlab\endcsname\relax\def\natexlab#1{#1}\fi

\bibitem[{{Abdalla} {et~al.}(2022){Abdalla}, {Abell{\'a}n}, {Aboubrahim}, {Agnello}, {Akarsu}, {et~al.}}]{2022JHEAp..34...49A}
{Abdalla}, E., {Abell{\'a}n}, G.~F., {Aboubrahim}, A., {et~al.} 2022, Journal of High Energy Astrophysics, 34, 49

\bibitem[{{Aizpuru} {et~al.}(2021){Aizpuru}, {Arjona}, \& {Nesseris}}]{2021PhRvD.104d3521A}
{Aizpuru}, A., {Arjona}, R., \& {Nesseris}, S. 2021, \prd, 104, 043521

\bibitem[{{Alam} {et~al.}(2021){Alam}, {Aubert}, {Avila}, {Balland}, {Bautista}, {Bershady}, {Bizyaev}, {Blanton}, {Bolton}, {Bovy}, {Brinkmann}, {Brownstein}, {Burtin}, {Chabanier}, {Chapman}, {Choi}, {Chuang}, {Comparat}, {Cousinou}, {Cuceu}, {Dawson}, {de la Torre}, {de Mattia}, {Agathe}, {des Bourboux}, {Escoffier}, {Etourneau}, {Farr}, {Font-Ribera}, {Frinchaboy}, {Fromenteau}, {Gil-Mar{\'\i}n}, {Le Goff}, {Gonzalez-Morales}, {Gonzalez-Perez}, {Grabowski}, {Guy}, {Hawken}, {Hou}, {Kong}, {Parker}, {Klaene}, {Kneib}, {Lin}, {Long}, {Lyke}, {de la Macorra}, {Martini}, {Masters}, {Mohammad}, {Moon}, {Mueller}, {Mu{\~n}oz-Guti{\'e}rrez}, {Myers}, {Nadathur}, {Neveux}, {Newman}, {Noterdaeme}, {Oravetz}, {Oravetz}, {Palanque-Delabrouille}, {Pan}, {Paviot}, {Percival}, {P{\'e}rez-R{\`a}fols}, {Petitjean}, {Pieri}, {Prakash}, {Raichoor}, {Ravoux}, {Rezaie}, {Rich}, {Ross}, {Rossi}, {Ruggeri}, {Ruhlmann-Kleider}, {S{\'a}nchez}, {S{\'a}nchez}, {S{\'a}nchez-Gallego}, {Sayres}, {Schneider}, {Seo}, {Shafieloo},
  {Slosar}, {Smith}, {Stermer}, {Tamone}, {Tinker}, {Tojeiro}, {Vargas-Maga{\~n}a}, {Variu}, {Wang}, {Weaver}, {Weijmans}, {Y{\`e}che}, {Zarrouk}, {Zhao}, {Zhao}, \& {Zheng}}]{2021PhRvD.103h3533A}
{Alam}, S., {Aubert}, M., {Avila}, S., {et~al.} 2021, \prd, 103, 083533

\bibitem[{{Alfano} {et~al.}(2024{\natexlab{a}}){Alfano}, {Capozziello}, {Luongo}, \& {Muccino}}]{2024JHEAp..42..178A}
{Alfano}, A.~C., {Capozziello}, S., {Luongo}, O., \& {Muccino}, M. 2024{\natexlab{a}}, Journal of High Energy Astrophysics, 42, 178

\bibitem[{{Alfano} {et~al.}(2024{\natexlab{b}}){Alfano}, {Luongo}, \& {Muccino}}]{2024A&A...686A..30A}
{Alfano}, A.~C., {Luongo}, O., \& {Muccino}, M. 2024{\natexlab{b}}, \aap, 686, A30

\bibitem[{{Alfano} {et~al.}(2024{\natexlab{c}}){Alfano}, {Luongo}, \& {Muccino}}]{2024arXiv240802536A}
{Alfano}, A.~C., {Luongo}, O., \& {Muccino}, M. 2024{\natexlab{c}}, arXiv e-prints, arXiv:2408.02536

\bibitem[{{Amati} {et~al.}(2019){Amati}, {D'Agostino}, {Luongo}, {Muccino}, \& {Tantalo}}]{2019MNRAS.486L..46A}
{Amati}, L., {D'Agostino}, R., {Luongo}, O., {Muccino}, M., \& {Tantalo}, M. 2019, \mnras, 486, L46

\bibitem[{{Arjona} {et~al.}(2019){Arjona}, {Cardona}, \& {Nesseris}}]{2019PhRvD..99d3516A}
{Arjona}, R., {Cardona}, W., \& {Nesseris}, S. 2019, \prd, 99, 043516

\bibitem[{{Aubourg} {et~al.}(2015){Aubourg}, {Bailey}, {Bautista}, {Beutler}, {Bhardwaj}, {et~al.}}]{2015PhRvD..92l3516A}
{Aubourg}, {\'E}., {Bailey}, S., {Bautista}, J.~E., {et~al.} 2015, \prd, 92, 123516

\bibitem[{{Belfiglio} {et~al.}(2022){Belfiglio}, {Giamb{\`o}}, \& {Luongo}}]{mio2022}
{Belfiglio}, A., {Giamb{\`o}}, R., \& {Luongo}, O. 2022, arXiv e-prints, arXiv:2206.14158

\bibitem[{Bernal \& Libanore(2023)}]{bernalcosmic}
Bernal, J.~L. \& Libanore, S. 2023, Cosmic Tensions -- Lecture Notes

\bibitem[{{Beutler} {et~al.}(2011){Beutler}, {Blake}, {Colless}, {Jones}, {Staveley-Smith}, {et~al.}}]{2011MNRAS.416.3017B}
{Beutler}, F., {Blake}, C., {Colless}, M., {et~al.} 2011, \mnras, 416, 3017

\bibitem[{{Blake} {et~al.}(2012){Blake}, {Brough}, {Colless}, {Contreras}, {Couch}, {et~al.}}]{2012MNRAS.425..405B}
{Blake}, C., {Brough}, S., {Colless}, M., {et~al.} 2012, \mnras, 425, 405

\bibitem[{{Bonamente} {et~al.}(2006){Bonamente}, {Joy}, {LaRoque}, {Carlstrom}, {Reese}, \& {Dawson}}]{2006ApJ...647...25B}
{Bonamente}, M., {Joy}, M.~K., {LaRoque}, S.~J., {et~al.} 2006, \apj, 647, 25

\bibitem[{{Borghi} {et~al.}(2022){Borghi}, {Moresco}, \& {Cimatti}}]{2022ApJ...928L...4B}
{Borghi}, N., {Moresco}, M., \& {Cimatti}, A. 2022, \apjl, 928, L4

\bibitem[{{Capozziello} {et~al.}(2013){Capozziello}, {De Laurentis}, {Luongo}, \& {Ruggeri}}]{2013Galax...1..216C}
{Capozziello}, S., {De Laurentis}, M., {Luongo}, O., \& {Ruggeri}, A. 2013, Galaxies, 1, 216

\bibitem[{{Carloni} {et~al.}(2024){Carloni}, {Luongo}, \& {Muccino}}]{2024arXiv240412068C}
{Carloni}, Y., {Luongo}, O., \& {Muccino}, M. 2024, arXiv e-prints, arXiv:2404.12068

\bibitem[{{Carlstrom} {et~al.}(2002){Carlstrom}, {Holder}, \& {Reese}}]{2002ARA&A..40..643C}
{Carlstrom}, J.~E., {Holder}, G.~P., \& {Reese}, E.~D. 2002, \araa, 40, 643

\bibitem[{{Carroll}(2001)}]{2001LRR.....4....1C}
{Carroll}, S.~M. 2001, Living Reviews in Relativity, 4, 1

\bibitem[{{Colg{\'a}in} {et~al.}(2024){Colg{\'a}in}, {Dainotti}, {Capozziello}, {Pourojaghi}, {Sheikh-Jabbari}, \& {Stojkovic}}]{2024arXiv240408633C}
{Colg{\'a}in}, E.~{\'O}., {Dainotti}, M.~G., {Capozziello}, S., {et~al.} 2024, arXiv e-prints, arXiv:2404.08633

\bibitem[{{Conley} {et~al.}(2011){Conley}, {Guy}, {Sullivan}, {Regnault}, {Astier}, {Balland}, {Basa}, {Carlberg}, {Fouchez}, {Hardin}, {Hook}, {Howell}, {Pain}, {Palanque-Delabrouille}, {Perrett}, {Pritchet}, {Rich}, {Ruhlmann-Kleider}, {Balam}, {Baumont}, {Ellis}, {Fabbro}, {Fakhouri}, {Fourmanoit}, {Gonz{\'a}lez-Gait{\'a}n}, {Graham}, {Hudson}, {Hsiao}, {Kronborg}, {Lidman}, {Mourao}, {Neill}, {Perlmutter}, {Ripoche}, {Suzuki}, \& {Walker}}]{2011ApJS..192....1C}
{Conley}, A., {Guy}, J., {Sullivan}, M., {et~al.} 2011, \apjs, 192, 1

\bibitem[{{Copeland} {et~al.}(2006){Copeland}, {Sami}, \& {Tsujikawa}}]{2006IJMPD..15.1753C}
{Copeland}, E.~J., {Sami}, M., \& {Tsujikawa}, S. 2006, International Journal of Modern Physics D, 15, 1753

\bibitem[{{Cuceu} {et~al.}(2019){Cuceu}, {Farr}, {Lemos}, \& {Font-Ribera}}]{2019JCAP...10..044C}
{Cuceu}, A., {Farr}, J., {Lemos}, P., \& {Font-Ribera}, A. 2019, \jcap, 2019, 044

\bibitem[{{D'Agostino} {et~al.}(2022){D'Agostino}, {Luongo}, \& {Muccino}}]{2022CQGra..39s5014D}
{D'Agostino}, R., {Luongo}, O., \& {Muccino}, M. 2022, Classical and Quantum Gravity, 39, 195014

\bibitem[{{De Filippis} {et~al.}(2005){De Filippis}, {Sereno}, {Bautz}, \& {Longo}}]{2005ApJ...625..108D}
{De Filippis}, E., {Sereno}, M., {Bautz}, M.~W., \& {Longo}, G. 2005, \apj, 625, 108

\bibitem[{{DESI Collaboration}(2024)}]{2024arXiv240403002D}
{DESI Collaboration}. 2024, arXiv e-prints, arXiv:2404.03002

\bibitem[{{Di Valentino} {et~al.}(2021){Di Valentino}, {Mena}, {Pan}, {Visinelli}, {Yang}, {et~al.}}]{2021CQGra..38o3001D}
{Di Valentino}, E., {Mena}, O., {Pan}, S., {et~al.} 2021, Classical and Quantum Gravity, 38, 153001

\bibitem[{{Dunsby} \& {Luongo}(2016)}]{2016IJGMM..1330002D}
{Dunsby}, P. K.~S. \& {Luongo}, O. 2016, International Journal of Geometric Methods in Modern Physics, 13, 1630002

\bibitem[{{Efstathiou} \& {Bond}(1999)}]{1999MNRAS.304...75E}
{Efstathiou}, G. \& {Bond}, J.~R. 1999, \mnras, 304, 75

\bibitem[{{Giar{\`e}} {et~al.}(2024){Giar{\`e}}, {Najafi}, {Pan}, {Di Valentino}, \& {Firouzjaee}}]{2024JCAP...10..035G}
{Giar{\`e}}, W., {Najafi}, M., {Pan}, S., {Di Valentino}, E., \& {Firouzjaee}, J.~T. 2024, \jcap, 2024, 035

\bibitem[{{Glanville} {et~al.}(2021){Glanville}, {Howlett}, \& {Davis}}]{2021MNRAS.503.3510G}
{Glanville}, A., {Howlett}, C., \& {Davis}, T.~M. 2021, \mnras, 503, 3510

\bibitem[{{Haridasu} {et~al.}(2018){Haridasu}, {Lukovi{\'c}}, {Moresco}, \& {Vittorio}}]{2018JCAP...10..015H}
{Haridasu}, B.~S., {Lukovi{\'c}}, V.~V., {Moresco}, M., \& {Vittorio}, N. 2018, \jcap, 2018, 015

\bibitem[{{Hastings}(1970)}]{1970Bimka..57...97H}
{Hastings}, W.~K. 1970, Biometrika, 57, 97

\bibitem[{{Jiao} {et~al.}(2023){Jiao}, {Borghi}, {Moresco}, \& {Zhang}}]{2023ApJS..265...48J}
{Jiao}, K., {Borghi}, N., {Moresco}, M., \& {Zhang}, T.-J. 2023, \apjs, 265, 48

\bibitem[{{Jimenez} \& {Loeb}(2002{\natexlab{a}})}]{2002ApJ...573...37J}
{Jimenez}, R. \& {Loeb}, A. 2002{\natexlab{a}}, \apj, 573, 37

\bibitem[{{Jimenez} \& {Loeb}(2002{\natexlab{b}})}]{Jimenez2002}
{Jimenez}, R. \& {Loeb}, A. 2002{\natexlab{b}}, \apj, 573, 37

\bibitem[{{Luongo} \& {Muccino}(2018)}]{nostro}
{Luongo}, O. \& {Muccino}, M. 2018, \prd, 98, 103520

\bibitem[{{Luongo} \& {Muccino}(2021{\natexlab{a}})}]{2021Galax...9...77L}
{Luongo}, O. \& {Muccino}, M. 2021{\natexlab{a}}, Galaxies, 9, 77

\bibitem[{{Luongo} \& {Muccino}(2021{\natexlab{b}})}]{2021MNRAS.503.4581L}
{Luongo}, O. \& {Muccino}, M. 2021{\natexlab{b}}, \mnras, 503, 4581

\bibitem[{{Luongo} \& {Muccino}(2023)}]{2023MNRAS.518.2247L}
{Luongo}, O. \& {Muccino}, M. 2023, \mnras, 518, 2247

\bibitem[{{Luongo} \& {Muccino}(2024)}]{2024A&A...690A..40L}
{Luongo}, O. \& {Muccino}, M. 2024, \aap, 690, A40

\bibitem[{{Metropolis} {et~al.}(1953){Metropolis}, {Rosenbluth}, {Rosenbluth}, {Teller}, \& {Teller}}]{1953JChPh..21.1087M}
{Metropolis}, N., {Rosenbluth}, A.~W., {Rosenbluth}, M.~N., {Teller}, A.~H., \& {Teller}, E. 1953, \jcp, 21, 1087

\bibitem[{{Montiel} {et~al.}(2021){Montiel}, {Cabrera}, \& {Hidalgo}}]{2021MNRAS.501.3515M}
{Montiel}, A., {Cabrera}, J.~I., \& {Hidalgo}, J.~C. 2021, \mnras, 501, 3515

\bibitem[{{Moresco}(2015)}]{Moresco2015}
{Moresco}, M. 2015, \mnras, 450, L16

\bibitem[{{Moresco} {et~al.}(2022){Moresco}, {Amati}, {Amendola}, {Birrer}, {Blakeslee}, {Cantiello}, {Cimatti}, {Darling}, {Della Valle}, {Fishbach}, {Grillo}, {Hamaus}, {Holz}, {Izzo}, {Jimenez}, {Lusso}, {Meneghetti}, {Piedipalumbo}, {Pisani}, {Pourtsidou}, {Pozzetti}, {Quartin}, {Risaliti}, {Rosati}, \& {Verde}}]{2022LRR....25....6M}
{Moresco}, M., {Amati}, L., {Amendola}, L., {et~al.} 2022, Living Reviews in Relativity, 25, 6

\bibitem[{{Moresco} {et~al.}(2012){Moresco}, {Cimatti}, {Jimenez}, {Pozzetti}, {Zamorani}, {et~al.}}]{Moresco2012}
{Moresco}, M., {Cimatti}, A., {Jimenez}, R., {et~al.} 2012, \jcap, 2012, 006

\bibitem[{{Moresco} {et~al.}(2016){Moresco}, {Pozzetti}, {Cimatti}, {Jimenez}, {Maraston}, {et~al.}}]{Moresco2016}
{Moresco}, M., {Pozzetti}, L., {Cimatti}, A., {et~al.} 2016, \jcap, 2016, 014

\bibitem[{{Muccino} {et~al.}(2023){Muccino}, {Luongo}, \& {Jain}}]{2023MNRAS.523.4938M}
{Muccino}, M., {Luongo}, O., \& {Jain}, D. 2023, \mnras, 523, 4938

\bibitem[{{Peebles} \& {Ratra}(2003)}]{2003RvMP...75..559P}
{Peebles}, P.~J. \& {Ratra}, B. 2003, Reviews of Modern Physics, 75, 559

\bibitem[{{Perlmutter} {et~al.}(1999){Perlmutter}, {Aldering}, {Goldhaber}, {Knop}, {Nugent}, {et~al.}}]{1999ApJ...517..565P}
{Perlmutter}, S., {Aldering}, G., {Goldhaber}, G., {et~al.} 1999, \apj, 517, 565

\bibitem[{{Planck Collaboration}(2020)}]{Planck2018}
{Planck Collaboration}. 2020, A\&A, 641, A6

\bibitem[{{Ratsimbazafy} {et~al.}(2017){Ratsimbazafy}, {Loubser}, {Crawford}, {Cress}, {Bassett}, {et~al.}}]{2017MNRAS.467.3239R}
{Ratsimbazafy}, A.~L., {Loubser}, S.~I., {Crawford}, S.~M., {et~al.} 2017, \mnras, 467, 3239

\bibitem[{{Riess} {et~al.}(1998){Riess}, {Filippenko}, {Challis}, {Clocchiatti}, {Diercks}, {et~al.}}]{1998AJ....116.1009R}
{Riess}, A.~G., {Filippenko}, A.~V., {Challis}, P., {et~al.} 1998, \aj, 116, 1009

\bibitem[{{Riess} {et~al.}(2022){Riess}, {Yuan}, {Macri}, {Scolnic}, , {Brout}, {et~al.}}]{2022ApJ...934L...7R}
{Riess}, A.~G., {Yuan}, W., {Macri}, L.~M., {et~al.} 2022, \apjl, 934, L7

\bibitem[{{Risaliti} \& {Lusso}(2019)}]{2019NatAs...3..272R}
{Risaliti}, G. \& {Lusso}, E. 2019, Nature Astronomy, 3, 272

\bibitem[{{Sch{\"o}neberg}(2024)}]{2024JCAP...06..006S}
{Sch{\"o}neberg}, N. 2024, \jcap, 2024, 006

\bibitem[{{Scolnic} {et~al.}(2022){Scolnic}, {Brout}, {Carr}, {Riess}, {Davis}, {Dwomoh}, {Jones}, {Ali}, {Charvu}, {Chen}, {Peterson}, {Popovic}, {Rose}, {Wood}, {Brown}, {Chambers}, {Coulter}, {Dettman}, {Dimitriadis}, {Filippenko}, {Foley}, {Jha}, {Kilpatrick}, {Kirshner}, {Pan}, {Rest}, {Rojas-Bravo}, {Siebert}, {Stahl}, \& {Zheng}}]{2022ApJ...938..113S}
{Scolnic}, D., {Brout}, D., {Carr}, A., {et~al.} 2022, \apj, 938, 113

\bibitem[{{Shafieloo}(2007)}]{2007MNRAS.380.1573S}
{Shafieloo}, A. 2007, \mnras, 380, 1573

\bibitem[{{Shafieloo} \& {Clarkson}(2010)}]{2010PhRvD..81h3537S}
{Shafieloo}, A. \& {Clarkson}, C. 2010, \prd, 81, 083537

\bibitem[{{Simon} {et~al.}(2005){Simon}, {Verde}, \& {Jimenez}}]{Simon2005}
{Simon}, J., {Verde}, L., \& {Jimenez}, R. 2005, \prd, 71, 123001

\bibitem[{{Stern} {et~al.}(2010){Stern}, {Jimenez}, {Verde}, {Kamionkowski}, \& {Stanford}}]{Stern2010}
{Stern}, D., {Jimenez}, R., {Verde}, L., {Kamionkowski}, M., \& {Stanford}, S.~A. 2010, \jcap, 2010, 008

\bibitem[{{Sunyaev} \& {Zeldovich}(1970)}]{1970CoASP...2...66S}
{Sunyaev}, R.~A. \& {Zeldovich}, Y.~B. 1970, Comments on Astrophysics and Space Physics, 2, 66

\bibitem[{{Sunyaev} \& {Zeldovich}(1972)}]{1972CoASP...4..173S}
{Sunyaev}, R.~A. \& {Zeldovich}, Y.~B. 1972, Comments on Astrophysics and Space Physics, 4, 173

\bibitem[{{Tomasetti} {et~al.}(2023){Tomasetti}, {Moresco}, {Borghi}, {Jiao}, {Cimatti}, {Pozzetti}, {Carnall}, {McLure}, \& {Pentericci}}]{Tomasetti2023}
{Tomasetti}, E., {Moresco}, M., {Borghi}, N., {et~al.} 2023, \aap, 679, A96

\bibitem[{{Wang}(2024)}]{2024arXiv240413833W}
{Wang}, D. 2024, arXiv e-prints, arXiv:2404.13833

\bibitem[{{Weinberg}(2008)}]{2008cosm.book.....W}
{Weinberg}, S. 2008, {Cosmology}

\bibitem[{{Wolf} \& {Ferreira}(2023)}]{2023PhRvD.108j3519W}
{Wolf}, W.~J. \& {Ferreira}, P.~G. 2023, \prd, 108, 103519

\bibitem[{{Wolf} {et~al.}(2024{\natexlab{a}}){Wolf}, {Ferreira}, \& {Garc{\'\i}a-Garc{\'\i}a}}]{2024arXiv240917019W}
{Wolf}, W.~J., {Ferreira}, P.~G., \& {Garc{\'\i}a-Garc{\'\i}a}, C. 2024{\natexlab{a}}, arXiv e-prints, arXiv:2409.17019

\bibitem[{{Wolf} {et~al.}(2024{\natexlab{b}}){Wolf}, {Garc{\'\i}a-Garc{\'\i}a}, {Bartlett}, \& {Ferreira}}]{2024PhRvD.110h3528W}
{Wolf}, W.~J., {Garc{\'\i}a-Garc{\'\i}a}, C., {Bartlett}, D.~J., \& {Ferreira}, P.~G. 2024{\natexlab{b}}, \prd, 110, 083528

\bibitem[{{Zhang} {et~al.}(2014){Zhang}, {Zhang}, {Yuan}, {Liu}, {Zhang}, {et~al.}}]{Zhang2014}
{Zhang}, C., {Zhang}, H., {Yuan}, S., {et~al.} 2014, Research in Astronomy and Astrophysics, 14, 1221

\end{thebibliography}

\begin{appendix}
\section{MCMC1 and MCMC2 details}\label{app:A}

We used a modified version of the code from \citet{2019PhRvD..99d3516A}, which is based on the Metropolis-Hastings algorithm \citep{1953JChPh..21.1087M, 1970Bimka..57...97H}.

For each of the MCMC analyses run in this paper:
\begin{itemize}
    \item[-] we worked out a preliminary MCMC analysis to obtain a test covariance matrix, and then
    \item[-] we performed the actual MCMC analysis to produce a single chain.
\end{itemize}
For both MCMC1 and MCMC2 chains the initial $100$ steps have been removed as burn-in, leaving chains with overall lengths of $N\sim1.3\times10^4$.
To asses their convergence we computed the corresponding autocorrelation functions  at lag $k$
\begin{equation}
\label{autocorrelation}
\rho_k(\mathcal L)=\frac{\sum^{N-k}_{t=1}\left(\mathcal L_t-\bar{\mathcal L}\right)\left(\mathcal L_{t+k}-\bar{\mathcal L}\right)}{\sum^{N}_{t=1}\left(\mathcal L_t-\bar{\mathcal L}\right)^2}\,,
\end{equation}
where $\mathcal L_t$ is the value of the log-likelihood at the step $t$ and $\bar{\mathcal L}$ is the mean of the chain.
From Eq.~\eqref{autocorrelation}, we define the autocorrelation length $l$ as the lag beyond which the ACF drops below the threshold $\rho_l(\mathcal L)=0.01$.
MCMC1 and MCM2 analyses provide small values $l=\{12,\,13\}$, respectively, indicating that most of the samples in the chains are independent, as indicated by the effective sample size
\begin{equation}
\label{EES}
{\rm ESS} = \frac{N}{1+2\sum^{l}_{k=1}\rho_k(\mathcal L)}\,,
\end{equation}
giving high values ${\rm ESS}=\{2178,\,1967\}$, respectively.

\begin{figure*}
\centering
\includegraphics[width=.99\hsize,clip]{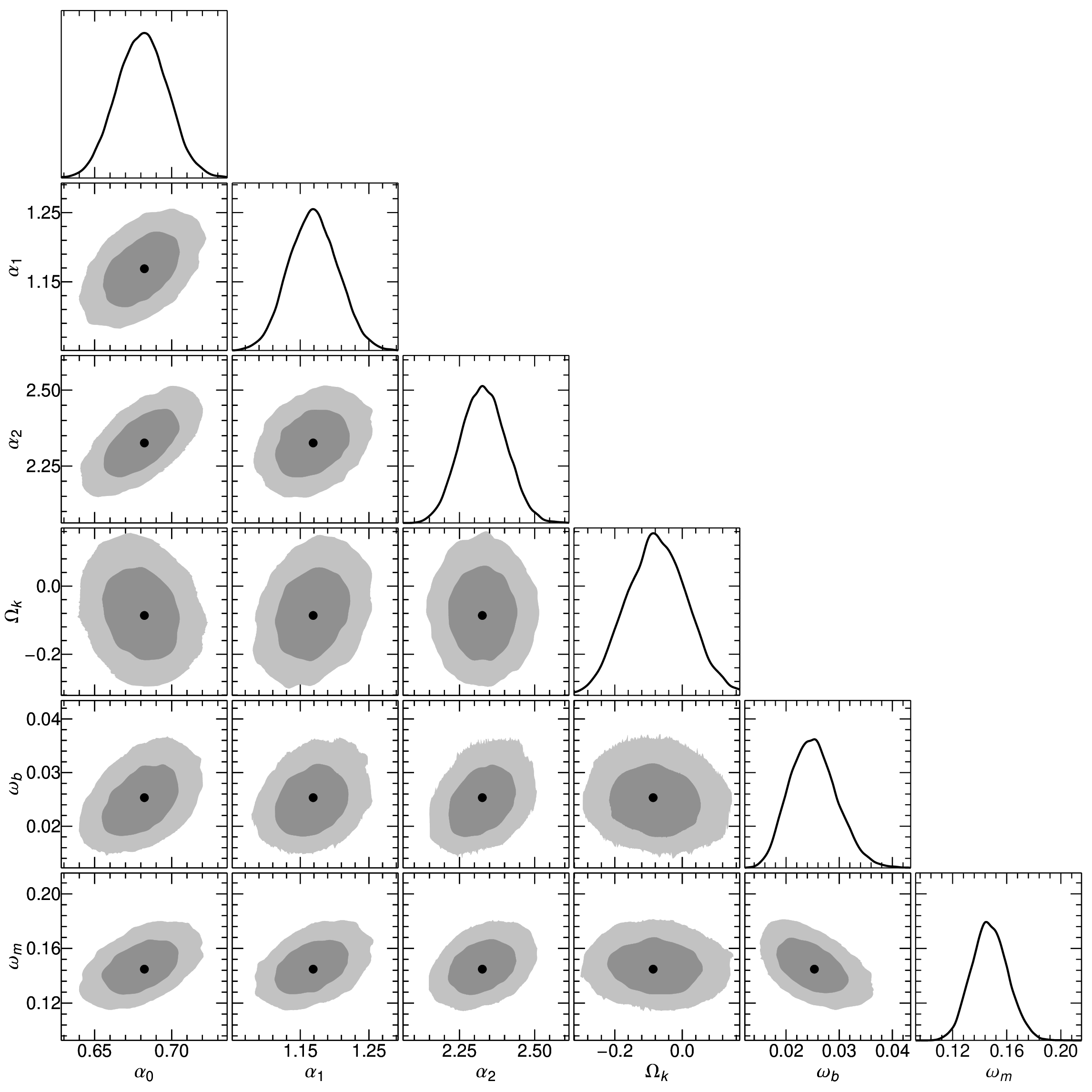}
\caption{MCMC1 contour plots. Darker (lighter) areas display $1$--$\sigma$ ($2$--$\sigma$) confidence regions.}
\label{fig:cont1}
\end{figure*}

\begin{figure*}
\centering
\includegraphics[width=.99\hsize,clip]{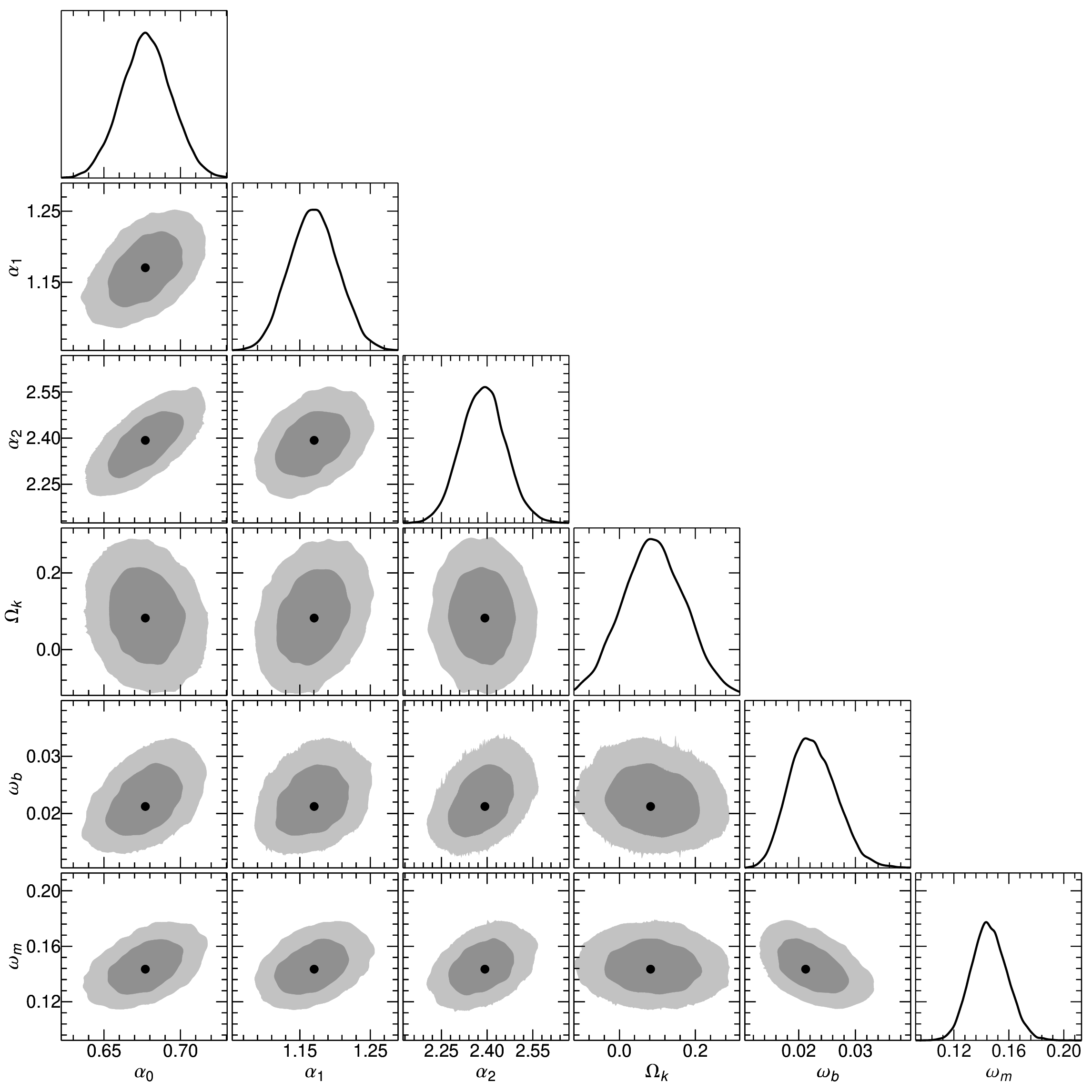}
\caption{MCMC2 contour plots. Darker (lighter) areas dispaly $1$--$\sigma$ ($2$--$\sigma$) confidence regions.}
\label{fig:cont2}
\end{figure*}

The best-fit parameters got from MCMC1 and MCMC2 analyses are portrayed in the $1$--$\sigma$ and $2$--$\sigma$ contour plots of Figs.~\ref{fig:cont1}--\ref{fig:cont2}.

\section{Coefficients of the dark energy reconstruction}\label{app:B}

We here show the explicit expressions of the coefficients $\beta_i$ used throughout the DE reconstruction reported in Eq.~\eqref{DEz}:
\begin{subequations}
    \begin{align}
\beta_0 =&\,\frac{\left[\alpha_2+\alpha _0 \left(1+z_{\rm m}\right){}^2-2 \alpha_1 \left(1+z_{\rm m}\right)\right]^2}{\alpha_0^2 z_{\rm m}^4}\,,\\
\beta_1 =& -\frac{4\sqrt{\beta_0}\left[\alpha _2+\alpha _0 \left(1+z_{\rm m}\right)-\alpha_1 \left(2+z_{\rm m}\right)\right]}{\alpha _0 z_{\rm m}^2}\,,\\
\nonumber
\beta_2 =&\, \frac{6 \left(\alpha _0-2 \alpha _1+\alpha _2\right){}^2}{\alpha _0^2 z_{\rm m}^4} + \frac{12 \left(\alpha _0-\alpha _1\right) \left(\alpha _0-2 \alpha _1+\alpha _2\right)}{\alpha _0^2 z_{\rm m}^3} +\\
&\, \frac{2 \left(3 \alpha _0^2+\left(\alpha _2-6 \alpha _1\right) \alpha _0+2 \alpha _1^2\right)}{\alpha _0^2 z_{\rm m}^2}- \Omega_k\,,\\
\beta_3 =& \frac{\beta_1}{\sqrt{\beta_0}}\frac{\left(\alpha_0-2 \alpha _1+\alpha _2\right)}{\alpha_0 z_{\rm m}^2}-\frac{\omega_m}{\alpha_0^2}\,,\\
\beta_4 =&\,\frac{\left(\alpha _0-2 \alpha _1+\alpha _2\right)^2}{\alpha _0^2 z_{\rm m}^4}-\Omega_r\,.
    \end{align}
\end{subequations}
The errors $\sigma_{\beta_i}$ on the coefficients $\beta_i$ are computed using the covariance matrix $C^{\rm M}_{jk}$ of each MCMC analysis and the partial derivative matrixes $J_{ij}=\partial \beta_i/\partial x_j$ with variables $x_j=\{\alpha_0,\,\alpha_1,\,\alpha_2,\,\Omega_k,\,\omega_m\}$,
\begin{equation}
    \begin{cases}
    \displaystyle\sigma_{\beta_i}^2 = \sum_{j=0}^{4} \sum_{k=0}^{4} J_{ij}C^{\rm M}_{jk}J_{ik}, \quad & \quad i=0,...,3\,, \\
   \displaystyle\sigma_{\beta_i}^2 = \sum_{j=0}^{4} \sum_{k=0}^{4}  J_{ij}C^{\rm M}_{jk}J_{ik} + \sigma_{\Omega_r}^2, \quad & \quad i =4\,,
    \end{cases}
\end{equation}
where $\sigma_{\Omega_r}$ is the error on the radiation density parameter, which is not correlated with the variables $x_j$.

\end{appendix}

\end{document}